\documentclass{appolb}
\usepackage{epsfig}
\usepackage{graphicx}  

\begin{document}
\title{Some Intensive and Extensive Quantities in High-Energy Collisions
\thanks{Invited Talk at XXXI Max Born Symposium and HIC for FAIR Workshop, Three Days of Critical Behaviour in hot and dense QCD, Wroclaw-Poland, 14 - 16 JUNE 2013}
}
\author{A. Tawfik\thanks{http://atawfik.net/;~~~a.tawfik@eng.mti.edu.eg}
\address{Egyptian Center for Theoretical Physics (ECTP), MTI University, 11571 Cairo, Egypt and \\
World Laboratory for Cosmology And Particle Physics (WLCAPP), Cairo, Egypt}
}
\preprint{ECTP-2012-14\hspace*{0.5cm}and\hspace*{0.5cm}WLCAPP-2013-11}

\maketitle
\begin{abstract}
We review the evolution of some statistical and thermodynamical quantities measured in difference sizes of high-energy collisions at different energies. We differentiate between intensive and extensive quantities and discuss the importance of their distinguishability in characterizing possible critical phenomena of nuclear collisions at various energies with different initial conditions.
\end{abstract}

\section{Introduction}
The terminology "intensive and extensive quantity" was introduced by Richard C. Tolman \cite{tolman} in order to distinguish between different thermodynamical parameters,  properties,  variables, etc. Therefore, the defining of such quantities as intensive or extensive may depend on the way in which subsystems are arranged \cite{tolman}. In order to characterize possible critical phenomena of the nuclear collisions, which likely become complex at ultra high energy, various signatures have been proposed \cite{signatures}. It is obvious that the critical phenomena of intensive or extensive variables \cite{shuryak} should be differentiated. The extensive variables, like total charge multiplicity, obtain about equal contributions from the initial (due to fluctuations in spectators) and final stage (resonances).  The intensive variables, like particle ratios, are well described by resonances at the freeze-out \cite{Tawfik:2013eua,Tawfik:2013dba,Tawfik:2013pd,Tawfik:2005qn}. In the present work, we show how the distinguishability between extensive and intensive quantities behaves at various energies and with different initial conditions.

The implication of statistical-thermal models on high-energy physics dates back to about six decades \cite{hkoppe2}. Koppe introduced an almost-complete recipe for the statistical description of particle production \cite{tawfikKoppe}. The particle abundances in Fermi model \cite{fermi1} are treated by means of statistical weights. Furthermore,  Fermi model \cite{fermi1} gives a generalization of the {\it ''statistical model''}, in which one starts with a general cross-section formula and inserts into it a simplifying assumption about the matrix element of the process, which reflects many features of the high-energy reactions dominated by the density in phase space of the final states. In 1951, Pomeranchuk \cite{aa:pom} came up with the conjecture that a finite hadron size would imply a critical density above which the hadronic matter cannot be in the compound state, known as hadrons. Using all tools of statistical physics, Hagedorn introduced in 1965 the mass spectrum to describe the abundant formation of resonances with increasing masses and rotational degrees of freedom \cite{hgdrn} which relate the number of hadronic resonances to their masses as an exponential. Accordingly,  Hagedorn formulated the concept of limiting temperature based on the statistical bootstrap model. 

The statistical and thermodynamical variables, properties and parameters can be classified into intensive, extensive, normalized intensive and extensive, process and conjugate. There are physical properties which neither intensive nor extensive, e.g. electric resistance, invariant mass and special relativity. The intensivity is apparently additive and therefore a state variable. The intensive (bulk) properties do not depend on the system size or the amount of existing material. Therefore, it is scale invariant.  The extensivity is field and point variable but not additive. The extensive properties are additive for independent and non-interacting subsystems. They are directly proportional to the amount of existing material. Normalized intensive and extensive quantities are densities. They are not additive. The process depends on past history of the system. Therefore, they are differentiable, inexactly. The conjugates are intensive and extensive pairs, like temperature and entropy. For example, in grand canonical ensemble, strongly intensive quantities have been suggested as fluctuation measures not depending on the system volume and its fluctuations \cite{siq}. 
The charge distribution is inclusive, while isotropically resolved particle observation is an exclusive property. We review the evolution of some statistical and thermodynamical quantities measured in difference sizes of the high-energy collisions at different energies. 

The present paper is organized as follows. The intensivity and extensivity of statistical properties are shortly reviewed in section \ref{sec:inexStat}. The dissipative properties are elaborated in section \ref{sec:inexDiss}. The energy dependence of temperature shall be estimated in section \ref{sec:Temp}. Section \ref{sec:inexConc} is devoted to the conclusions and outlook.

\section{Statistical properties: multiplicities and particle ratios}
\label{sec:inexStat}

\begin{figure}[htb]
\begin{center}
\includegraphics[width=9.cm,height=6cm]{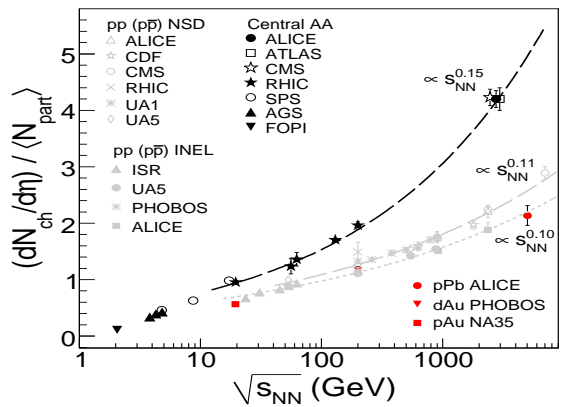}
\caption{A comparison of $dN_{ch}/d\eta$ per participating nucleon at
mid-rapidity in central heavy-ion collisions  to corresponding results from $p$+$p$($\bar{p}$) and $p(d)$+A collisions.
The quantities are given in physical units. Graph taken from Ref. \cite{epsl0tau}. }
\label{nFigjbrkEpslTau}  
\end{center} 
\end{figure}

Only two independent intensive variables are needed in order to fully specify the entire state of the system of interest. Other intensive properties can be derived from these known ones. An exclusive property implies that energy and momentum, for instance, of all products are measured. The intensivity means that some quantities of the products are left unmeasured.

An extensive comparison between the particle multiplicity $dN_{ch}/d\eta$ per participating nucleon at mid-rapidity in central heavy-ion collisions ~\cite{Back:2003hx,Abreu:2002fx,Adler:2001yq,Bearden:2001xw,Bearden:2001qq,Adcox:2000sp,Back:2000gw,Back:2002wb,Aamodt:2010pb,ATLAS:2011ag,Chatrchyan:2011pb} and the corresponding results from $p$+$p$($\bar{p}$)~\cite{ua1,ua5,Abelev:2008ab,Abe:1989td,Aamodt:2010pp,Khachatryan:2010xs,isr,Aamodt:2010ft,Alver:2010ck}  and $p(d)$+A collisions~\cite{ALICE:2012xs,Alber:1997sn,Back:2003hx} is presented in Fig. \ref{nFigjbrkEpslTau}. It is obvious that the energy dependence of the total multiplicity is distinguishable. In order words, the initial state plays an essential role. The extenstivity can be related to canonical ensemble, 
\begin{eqnarray}
Z(N,T,V) &=& \mathtt{Tr}_N\, \exp\left(-\frac{H}{T}\right),
\end{eqnarray}
where $H$ is the Hamiltonian, while grand canonical ensemble is related to intensivity,
\begin{eqnarray}
Z(\mu,T,V) &=& \mathtt{Tr}_N\, \exp\left(-\frac{H-\mu\, N}{T}\right),
\end{eqnarray}
where $N$ stands for the degrees of freedom. 
With Dirac delta function and when the chemical potential $\mu$ is Wick rotated, then extenstivity can be related to intensivity
\begin{eqnarray}
Z(N,T,V) &=& \frac{1}{2\, \pi} \int_0^{2\, \pi} Z(i T \theta,T,V)\, \exp(-i N \theta)\, d \theta.
\end{eqnarray}

\begin{figure}[thb]
\centering{
\includegraphics[width=5.45cm,height=8.cm,angle=-90]{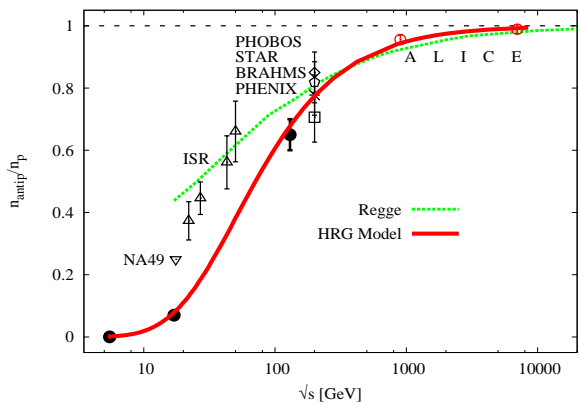}
\caption{$n_{\bar{p}}/n_p$ ratios depicted in whole available range of $\sqrt{s}$. Open symbols stand for the results from various $pp$ experiments (labeled). The solid symbols give the heavy-ion results from AGS, SPS and RHIC, respectively. The fitting of $pp$ results according to Regge model is given by the dashed curve \cite{alice2010}. The solid curve is the HRG results. Contrary to the dashed curve, the solid line is not a fitting to experimental data. The graph taken from Ref. \cite{Tawfik:2010pt}. } 
\label{fig:pap2010}
}
\end{figure}

In Fig. \ref{fig:pap2010}, the results of $\bar{p}/p$  calculated in HRG are represented by solid line, which  seems to be a kind of a universal curve. In heavy-ion collisions, the proton ratio varies strongly with the center-of-mass energy $\sqrt{s}$. The HRG models describes very well the heavy-ion results. Also, ALICE $pp$ results are reproduced by means of HRG model. The ratios from $pp$- and $AA$-collisions runs very close to unity implying almost vanishing matter-antimatter asymmetry. On the other hand, it can also be concluded that the statistical-thermal models including HRG seem to excellently describe the hadronization at very large energies and the condition deriving the chemical freeze-out at the final state of hadronization, the constant degrees of freedom or $S(\sqrt{s},T)=7 (4/\pi^2) V T^3$, seems to be valid at all center-of-mass-energies spanning between AGS and LHC.  So far, we conclude that the distinguishability between proton ratios in pp-collisions and that in AA-collisions disappears with increasing $\sqrt{s}$.

\section{Dissipative properties: elliptic flow}
\label{sec:inexDiss}

The azimuthal distribution with respect to the reaction plane reads
\begin{eqnarray}
\frac{d\, N}{d(\phi_i-\Psi_n)} &\sim& 1 + 2 \sum_{n=1} v_n\, \cos\left[n\left(\phi_i-\Psi_n\right)\right].
\end{eqnarray}
The reaction plane angle $\Psi_n$ is not directly measurable, but can be determined from particle azimuthal distributions. There are various possible sources of azimuthal correlations like, jet formation, resonances exist, which do not depend on the reaction plane (non-flow correlations). The Fourier coefficient $v_n$, which refers to the correlation in $n$ particle emission with respect to the reaction plane, is given by
\begin{eqnarray}
 v_n &=& \langle \cos\left[n\left(\phi_i-\Psi_n\right)\right]\rangle.
\end{eqnarray}

\begin{figure}[thb]
\centering{
\includegraphics[width=8cm,angle=-0]{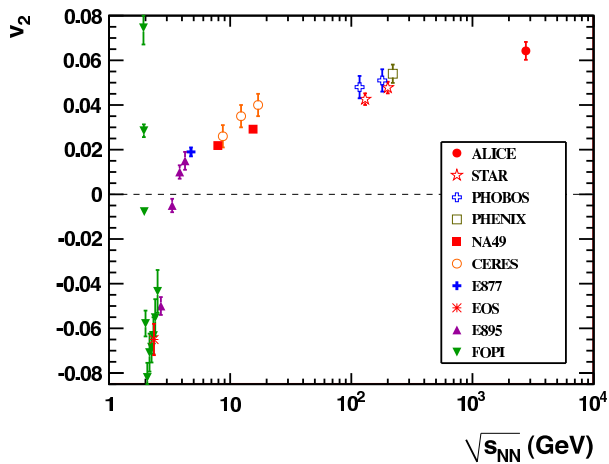}
\caption{Integrated elliptic flow measured in central heavy-ion collisions ($20-30\%$)  is given  in dependence on Nucleus-Nucleus center-of-mass energy. The graph taken from Ref. \cite{arnolt}. } 
\label{fig:pap2010v}
}
\end{figure}

Fig.~\ref{fig:pap2010v} shows data collected over about four decades spanning from GSI, AGS, SPS, RHIC to LHC facilities. The integrated elliptic flow measured in relative  central heavy-ion collisions ($20-30\%$)  is given  in dependence on Nucleus-Nucleus center-of-mass energy in Fig.~\ref{fig:pap2010v}.  For the comparison, the integrated elliptic flow is corrected for $p_{\rm t}$ cutoff of $0.2~$GeV/$c$. The estimated magnitude of this correction is  $12\pm 5\%$ based on calculations with Therminator. The figure shows that there is a continuous increase in the magnitude of elliptic flow for this centrality region
from RHIC to LHC energies. In comparison to the elliptic flow measurements in Au-Au collisions at $\sqrt{s_{NN}}=200$~GeV we observe about a $30\%$ increase in the magnitude of $v_2$ at $\sqrt{s_{NN}}=2.76$~TeV.  The rapid decrease of $v_2$ at very low energy, FOPI data, refers to {\it bounce-off}. Increasing $\sqrt{s_{NN}}$, a {\it squeeze-out} will set on. At larger energies, the behavior can be described by {\it in-plane elliptic flow}  due to pressure gradient.

the elliptic flow shows a rich structure; a transition from in-plane to out-of-plane and back to in-plane emission. Apparently, it is sensitive to the properties of the medium created in heavy-ion collisions. There are evidences that the elliptic flow of charged and identified particles indicates a strong rise of the expansion velocity of the medium (radial fow) at RHIC vs LHC.

On the other hand, it was assumed that there are no correlations due to elliptic flow in pp collisions at RHIC energy \cite{Eyyu}. The methods of measuring elliptic flow can hardly be employed with the currently available number of recorded pp interactions of ALICE at the LHC. Furthermore, none of available microscopic Monte Carlo (MC) models describes the development of  anisotropic flow in elementary hadron-hadron interactions yet \cite{Eyyu}. Particular non-perturbative approach was suggested as a mechanism of anisotropic flow might be a leading one in hadron collisions, since those have smaller geometrical extension and the probability of hydrodynamical generation of elliptic flow is lower compared to the collisions of nuclei \cite{troshin}.

$pp$ collisions simulated by PYTHIA, PHOJET and EPOS at $900$ and $7000~$GeV
are analyzed by two-particle correlation methods. The integrated $v_2$ coefficients reconstructed by the methods are found to vary from $10\%-15\%$. These values are attributed solely to the non-flow correlations \cite{Eyyu}.

\section{Hagedorn temperature: energy and system size dependence}
\label{sec:Temp}

\begin{figure}[thb]
\centering{
\includegraphics[width=6cm,angle=-0]{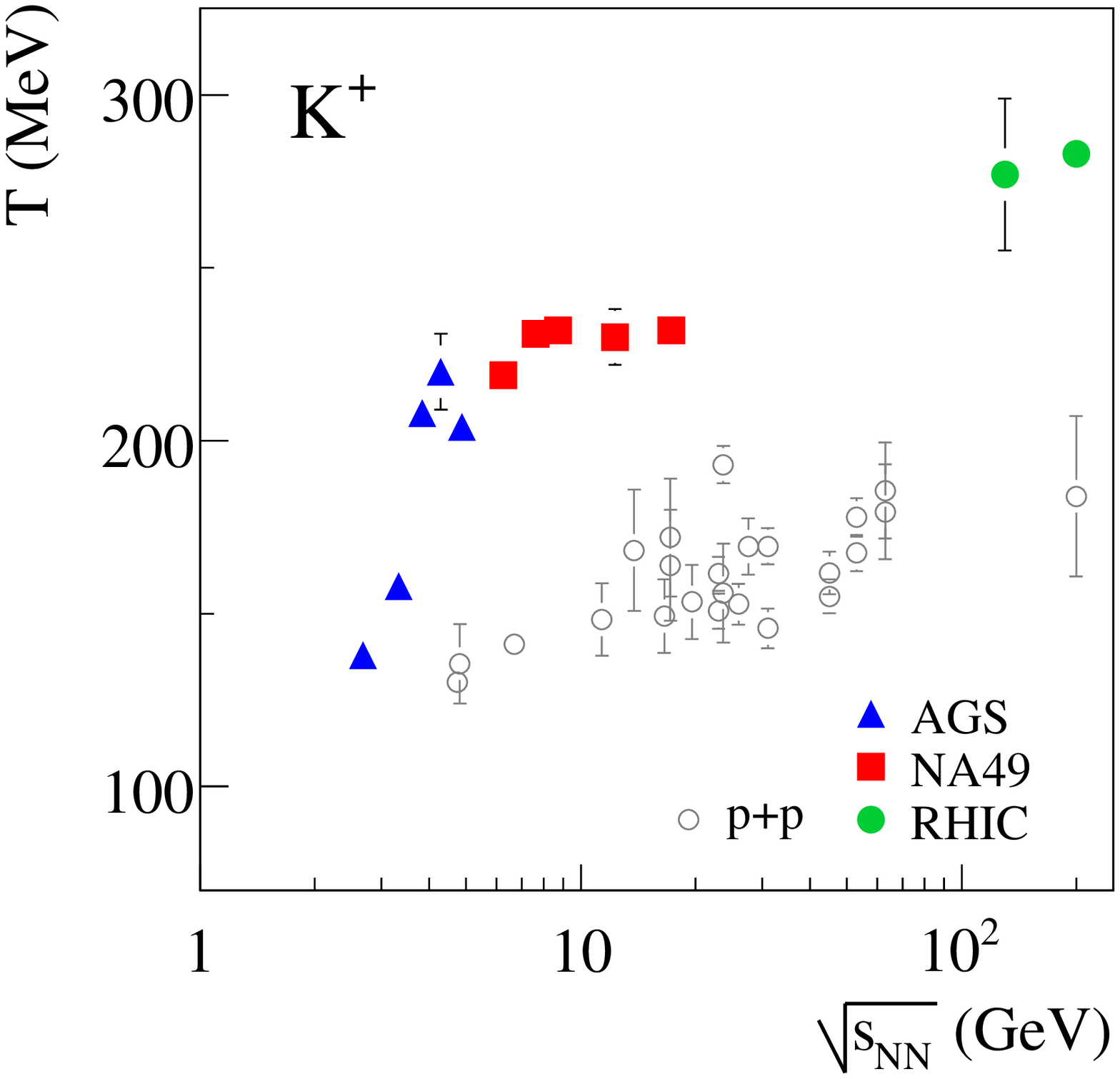}
\includegraphics[width=6cm,angle=-0]{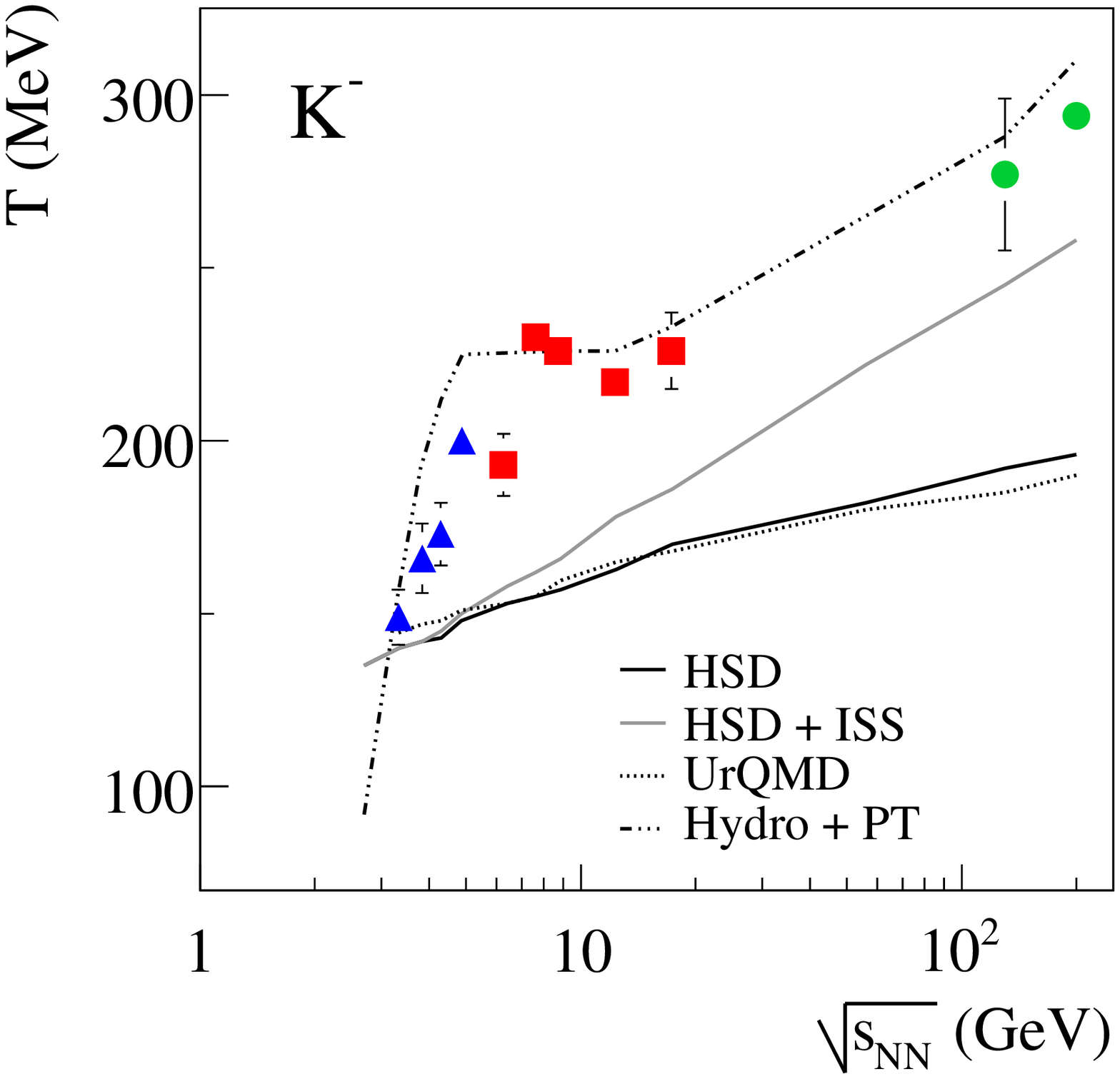}
\caption{Energy dependence of $T$ related to the transverse mass spectra of $K^+$ (left panel) and $K^-$ mesons (right panel) produced in central Pb+Pb and Au+Au collisions. The graphs taken from Ref. \cite{Alt}. } 
\label{fig:Tkpkm}
}
\end{figure}

The transverse mass spectra of well-identified particles have been studied at various energies, for instance \cite{Alt}.  Accordingly, Stefan-Boltzmann approximation results is
\begin{eqnarray}
\frac{1}{m_T}\, \frac{d N}{d m_T d y} &=& a \exp\left(-\frac{m_T}{T}\right),
\end{eqnarray}
where $m_T=\sqrt{p_T^2+m^2}$ is the dispersion relation and $a$ is a fitting parameter. 
Fig. \ref{fig:Tkpkm} presents the energy dependence of the inverse slope parameter $T$ of
the transverse mass spectra of $K^+$ (left panel) and $K^-$ mesons (right panel) produced in central Pb+Pb and Au+Au collisions.  There is a plateau at SPS energies \cite{Alt} which is preceded by a steep rise of $T$ measured at the AGS~\cite{AGS} and followed by an indication of  a further increase of the RHIC data~\cite{RHIC}.  Although the scatter of data points is large, $T$ appears to increase smoothly in $p+p(\bar{p})$ collisions \cite{kliemant},  left panel of Fig. \ref{fig:Tkpkm}. The dependence of $T$ on the system size is obvious. For completeness, we recall that the direct thermal photons have been used to estimate the Hagedorn temperature,
\begin{eqnarray}
T &=& \left\{
\begin{array}{lll} 
304\pm 51~\mathtt{MeV} & \mathtt{ALICE} & \cite{aliceT}\\
&&\\
221\pm19~\mathtt{MeV} & \mathtt{PHENIX} & \cite{phenixT}
\end{array}
 \right.
\end{eqnarray} 

The dependence on system size is illustrated in left panel of Fig. \ref{fig:Tkpkm}. The Hagedorn temperature in $pp$ collisions seems to be smaller than that in $AA$ collisions. Its variation with the center-of-mass energy is apparently weaker than the variation in $AA$ collisions. A much more systematic measurement would help in proving or disproving such a conclusion.

\section{Conclusions and outlook}
\label{sec:inexConc}

The ultimate goal of the physics program of high-energy collisions is the study of properties of strongly interacting matter under extreme conditions of temperature and/or compression. The particle multiplicities and their fluctuations and correlations are experimental tools to analyse the nature, composition, and size of the medium, from which they are originating. Of particular interest is the extent to which the measured particle yields are showing equilibration. Based on analysing the particle abundances or momentum spectra, the degree of equilibrium of the produced particles can be estimated. The particle abundances can help to establish the chemical composition of the system. The momentum spectra can give additional information on the dynamical evolution and the collective flow.

In order to characterize possible critical phenomena, signatures based on particle multiplicities and their fluctuations and correlations have been proposed. Intensive or extensive quantities should be separated, systematically. Extensivity obtains about equal contributions from the initial and final stage. Intensivity is well described by produced particles in final state. The present work introduces the importance of distinguishability between extensive and intensive quantities at various energies and in different system sizes.

\section*{Acknowledgement}
This work is financially supported by the World Laboratory for Cosmology And Particle Physics (WLCAPP), http://wlcapp.net/.

\end{document}